\documentclass[letterpaper,twocolumn,fleqn]{article} 

\usepackage{ist}
\usepackage{multirow}
\usepackage{multicol}

\usepackage[dvipsnames]{xcolor}
\usepackage{url}
\usepackage{cite}
\usepackage{hyperref}
\usepackage{subfigure}

\newcommand{\smalllskip}[0]{\vspace{1pt}}

\newcommand{\ie}[0]{\textit{i.e.},~}
\title{Quality Assessment of Super-Resolved Omnidirectional Image Quality Using Tangential Views}
\author{Cagri Ozcinar$^1$ and Aakanksha Rana$^2$\\Samsung R\&D$^1$, UK, Massachusetts Institute of Technology$^2$, USA}
\date{} 
\begin{document} 
\maketitle 
\thispagestyle{empty} 

\begin{abstract}
Omnidirectional images (ODIs), also known as 360-degree images, enable viewers to explore all directions of a given 360-degree scene from a fixed point. Designing an immersive imaging system with ODI is challenging as such systems require very large resolution coverage of the entire 360 viewing space to provide an enhanced quality of experience (QoE). Despite remarkable progress on single image super-resolution (SISR) methods with deep-learning techniques, no study for quality assessments of super-resolved ODIs exists to analyze the quality of such SISR techniques. This paper proposes an objective, full-reference quality assessment framework which studies quality measurement for ODIs generated by GAN-based and CNN-based SISR methods. The quality assessment framework offers to utilize tangential views to cope with the spherical nature of a given ODIs. The generated tangential views are distortion-free and can be efficiently scaled to high-resolution spherical data for SISR quality measurement. We extensively evaluate two state-of-the-art SISR methods using widely used full-reference SISR quality metrics adapted to our designed framework. In addition, our study reveals that most objective metric show high performance over CNN based SISR, while subjective tests favors GAN-based architectures.
\end{abstract}

\section{Introduction}
\label{sec:intro}

High-resolution omnidirectional images (ODIs) are quintessential to provide a sufficiently high-quality immersive experience for head-mounted displays. Single image super-resolution (SISR), a fundamentally ill-posed problem, enhances the resolution of a low-resolution image. While no existing consumer cameras can capture very large resolution, such as $\geq$8K, SISR techniques have emerged as a viable option for ODI \cite{360sr}. From traditional hand-crafted approaches to deep learning techniques, SISR methods for ODIs aim to construct the high-resolution ODIs using objective criteria adhering to SISR frameworks for standard images.

ODI processing systems can be optimized by leveraging the techniques of SISR. Similarly to the traditional image, quality assessment plays a vital role. However, no quality assessment framework exists in the literature that can measure and analyze the perceptual efficacy of the super-resolved ODIs while catering to its inherent spherical distortions. Without appropriate quality assessments, the viability of SISR for ODI techniques that promise immersive experience cannot be established.

Many recent learning-based SISR works have achieved remarkable improvements over traditional techniques. Majority of the existing SISR works aim to improve the super-resolved images' accuracy by minimizing the mean squared reconstruction error. With the invention of generative adversarial networks (GANs), state-of-the-art SISR methods claim to achieve higher perceptual quality and better realistic super-resolved images. 

While recent SISR works have improved quantitative results, there is an inconsistency between quantitative and qualitative results. Convolutional neural networks (CNNs) and generative adversarial networks (GANs) show great success for SISR. For instance, the super-resolution generative adversarial network (SRGAN)~\cite{ledig2017photo} shows that the high PSNR values do not always accord with better perceptual quality. The reason of this situation is that GAN-based SISR methods often produce realistic yet fake details and textures on the generated super-resolved images \cite{jinjin2020pipal}, bringing a challenge for existing image quality assessment methods.

SISR for ODI techniques should produce images with high perceptual quality to have an enhanced immersive viewing experience. However, the lack of understanding of the distortion of super-resolved ODIs motivates us to develop a quality measurement framework, and analyze the distortions of SISR and the performance of quality metrics. In particular, in our previous work \cite{360sr}, we showed that widely used quality metrics (SSIM, PSNR) developed for standard 2D images do not perform well for SISR on ODIs. Hence, we believe that analyzing quality metrics is crucial for designing an optimal SISR method for ODIs. Nevertheless, ODI quality evaluation analysis for SISR was not considered in the literature. To this end, in this study, we target the following objectives: 1) designing a novel SISR quality assessment framework for ODI; 2) performing a thorough evaluation study using state-of-the-art SISR methods for ODIs over state-of-the-art image quality assessment frameworks; 3) investigating the qualitative and quantitative results of CNN- and GAN- based SISR results.

This paper systematically evaluates a broad set of full-reference image quality assessment models with tangential views of ODI. We first developed an objective full-reference ODI quality assessment framework and studied quality measurement for ODIs generated by SISR methods. The proposed quality assessment framework considers the spherical nature of a given ODI with tangential views. These distortion-free tangential views are used to perform a large-scale comparison with eleven perceptual image quality assessment models and state-of-the-art SISR methods for ODI. This study presents the first quality assessment framework to assess the perceptual quality of super-resolved ODIs generated by state-of-the-art SISR methods. Moreover, this is the first evaluation study showcasing the performance of perceptual image quality assessment models on state-of-the-art SISR methods for ODIs. 

In this study, we present a straightforward comparison of objective and subjective analysis where we observe that GAN models shows superior performance in subjective tests over conventional CNN SISR architectures. However, contrary to it all the objective metrics favor the CNN models except a few. 

\section{Related Work}
\label{sec:related}
We first describe the latest contributions mostly related to our work, which are focused on: 1) quality assessment; 2) projections; 3) super-resolution. 

\subsection{Quality Assessment}

Traditional quality metrics for a standard 2D image do not have a high correlation with subjective scores because of inherent sphere-to-planar projection distortions of ODI~\cite{croci2019voronoi, roberto2020viewport}. Several PSNR-based methods have been proposed to cope with such projection distortions, such as weighted-to-spherical (WS-PSNR)~\cite{WS_PSNR}, spherical (S-PSNR)~\cite{S_PSNR}, and Craster parabolic projection (CPP-PSNR)~\cite{CPP_PSNR}. In WS-PSNR, the PSNR is applied at each pixel sample of the planar representation of ODI. Each distortion value on the planar is weighted by the area covered by the area on the sphere. In S-PSNR, the PSNR is computed for the sampling points uniformly distributed on a spherical surface. Furthermore, in CPP-PSNR, the PSNR is calculated in the Craster parabolic projection, characterized by low projection distortions. However, the previous related works are based on PSNR, which are not correlated well with subjective scores \cite{obj360, croci2020visual}. 

Although quality assessment techniques for ODI have been studied in recent years, the perceptual quality of ODI is still unexplored in image restoration techniques, such as super-resolution. In particular, most of the existing studies in this field consider only post-production~\cite{post360}, compression~\cite{tranquality}, and transmission artifacts~\cite{croci2019voronoi}.



\subsection{ODI Projections}
Sphere-to-planar projections inevitably introduce
Sphere-to-planar projections inevitably introduce
distortions which must be taken into account to estimate the ODI quality accurately. In particular, Sun~et al. \cite{sun2019mc360iqa} proposed a multi-channel CNN for blind ODI quality assessment. They generated six viewport images, representing six cubemap projection faces, to minimize the geometric distortions on the equirectangular images. However, ODIs with the equirectangular or cubemap image representations demonstrates severe image distortions~\cite{eder2019convolutions}. Also, \cite{croci2020visual} divides the ODV into a fixed number of planar patches based on the spherical Voronoi diagram. 

Instead, in this work, we consider 20-sided regular icoshedron which is the lowest distortion spherical representation~\cite{kimerling1999comparing}. In particular, our quality assessment framework is based on the tangent image spherical representation~\cite{eder2020tangent}. With this approach, we can subdivide a given ODI to match the spherical image resolution to the number of low-distortion patches. This approach provides a scalable solution for high-resolution ODIs.

\subsection{Super-resolution}

Thanks to the advanced learning capability\cite{ricip17,rbmvc,ricme17}, CNNs have demonstrated superior performance for the super-resolution task compared to the traditional approaches. For instance, an enhanced deep super-resolution network (EDSR)~\cite{lim2017enhanced} achieved the top PSNR rank in the NTIRE 2017 super-resolution challenge \cite{Timofte_2017_CVPR_Workshops}. This method modified the conventional ResNet architecture to the SISR task. The wide activation for efficient and accurate image super-resolution (WDSR)~\cite{yu2018wide} model demonstrated that expanding feature before activation leads to significant improvements for SISR. This network, WDSR, won the first places in NTIRE 2018 super-resolution challenge~\cite{timofte2018ntire} in three realistic tracks. With the invention of GANs, state-of-the-art SISR methods are providing better perceptual reconstruction results. In SISR, GANs provide a framework to generate plausible-looking natural images with perceptual quality. In particular, SRGAN~\cite{ledig2017photo} inspired many follow-up SISR works using GANs like ESRGAN~\cite{wang2018esrgan} and SFT-GAN~\cite{wang2018recovering}. 

We proposed an improved GAN-based SISR model for ODI and studied the problem of SISR in ODIs \cite{360sr}. In this work, we observed that quantitative quality measurements do not represent the visual super-resolved ODI results. 

Recent SISR methods claim to generate realistic textures and details incorporating feature loss and adversarial learning. However, they reveal their quantitative results with different full-reference quality metrics. Hence, Considering that there is a need to measure and understand super-resolved ODI quality, in this work, we performed subjective quality assessment experiments with diverse super-resolved ODIs, and extensive objective quality assessment using quality metrics on the tangential views.

\section{Objective Quality Assessment Framework}

\begin{figure*}
    \centering
    \includegraphics[width=0.8\linewidth]{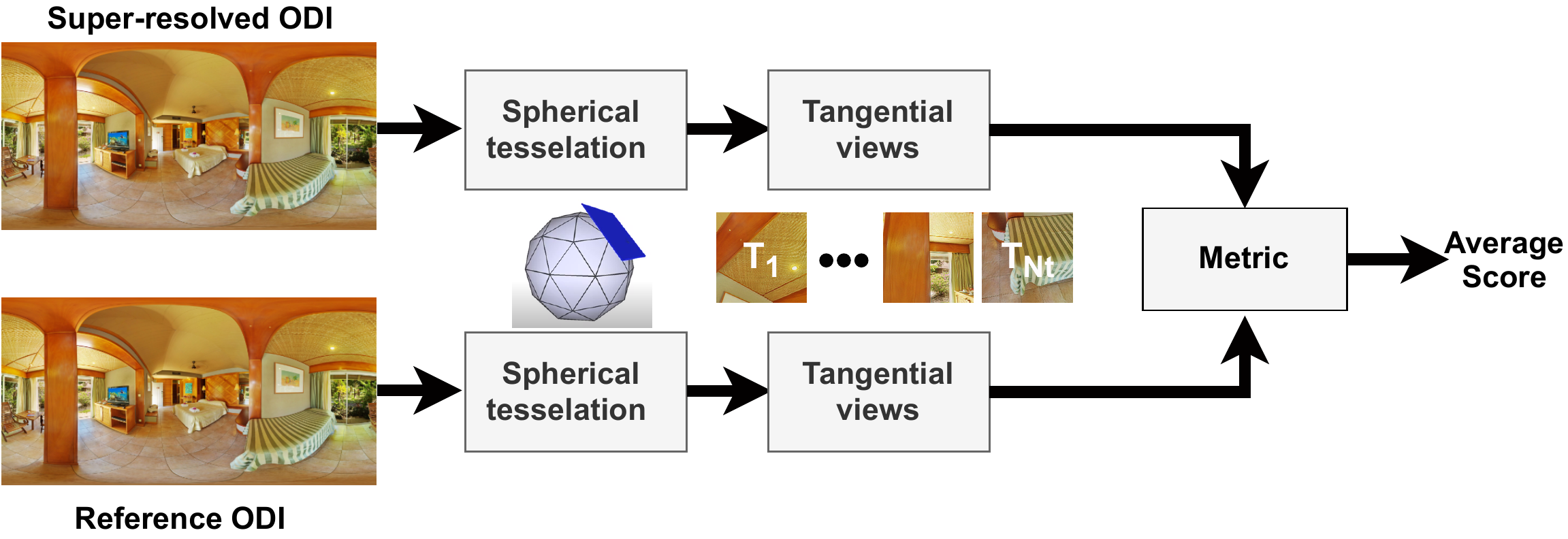}
    \caption{An overview of the proposed objective quality assessment framework for SISR.}
    \label{fig:pipeline}
     \vspace{-1.5em}
\end{figure*}

We consider the full-reference objective quality assessment framework to measure super-resolved ODIs, as depicted in Fig.~\ref{fig:pipeline}. Instead of learning a new model from scratch, our proposed framework allows us to use existing image quality assessment methods to ODI quality measurement. The proposed framework can take any ODI projection representations (e.g., equirectangular, cubemap, etc.) as super-resolved inputs and reference ODIs. We apply spherical tessellation with the icosahedron to subdivide the input ODIs. Tangential views are then extracted using the gnomonic projection as similar as~\cite{eder2020tangent}. We specifically focus on the full-reference quality metrics that take the human visual system's properties into account (e.g., SSIM, VIF, etc.). We apply the full-reference quality metrics developed for the traditional image to each tangential views. Finally, the metric scores for all tangential views are averaged to obtain the final objective quality score.

In the following sub-sections, we describes each component of the proposed objective quality assessment framework of super-resolved ODIs in detail. 

\subsection{Input Tangential Views}

Inspired by Eder et al~\cite{eder2020tangent}, we tessellate a sphere where every triangle of the icosahedron has exactly equal area to any desired resolution. For this, we render an image, called tangential view, onto an oriented pixel grids tangent to a subdivided icosahedron. With this approach, we can efficiently scale to represent high-resolution ODIs.

Our solution generates a number of low-distorted tangential views dependent on the resolution of the input ODI. The number of tangential views, $N_t$, is estimated by the faces of the base level of icosahedron as:

\begin{equation}
\centering
    N_t = f (4^b),
\end{equation}
where $f$ is a constant number representing the number of faces of the icosahedron, which is $20$. Also, $b$ represents a level of the subdivided icosahedron. In this paper, we propose to use $b=1$ which is the optimal number for many applications~\cite{eder2020tangent}. The spatial extent of this equation is a function of vertex resolution of the level $b-1$ icosahedron and the resolution of the image grid, as described in~\cite{eder2020tangent}.

\subsection{Quality measure using tangential views}~\label{proposed_tang}

The proposed framework extracts $N_t$ number of tangential views from both the super-resolved ODI and the reference ODI. The traditional full-reference quality metrics are then applied to each tangent views. Full-reference quality metrics compare a given distorted image to the complete reference image. The final objective score, t-metric, is then obtained by computing the average of the tangent images as following:
\begin{equation}
    \centering
     t-metric = \frac{\sum_{i=1}^{N_t}T_{i}}{N_t}, \quad with \quad T_i = Q(\tilde{t_i}, t_i),
\end{equation}
where $Q$ represent the quality score obtained by the full-reference quality metrics. Also, $\tilde{t_i}$ and $t_i$ are $i$-th tangential views of the super-resolved ODI $\tilde{y}$ and the complete reference image $y$, respectively.

We extend eleven traditional full-reference image quality metrics to ODI. We call tangential metrics, t-metric, to these extended objective quality metrics. The full-reference objective quality assessment methods used in this study is listed in Table.~\ref{tab:tmqi}. The selected quality metrics can be classified into four categories as:

\begin{itemize}
    \item Error visibility methods: These methods measure the distance between the distorted and the reference image pixels. In particular, the classical error estimation solution of mean square error ($l_2$-norm) and mean absolute error ($l_1$-norm) are widely used for optimizing the super-resolution algorithms. In this category, we utilized the most apparent distortion (MAD) metric~\cite{larson2010most} and normalized Laplacian pyramid (NLPD)~\cite{laparra2016perceptual}.

    \item Structural similarity measure: Structural similarity is used for measuring the similarity between the distorted and the reference image~\cite{wang2004image, wang2003multiscale}. These measurements extract three key features from the image, namely, luminance, contrast, and structure, and combined them for the final measure. In this category, we used structural similarity index (SSIM)~\cite{wang2004image}, multiscale (MS-SSIM)~\cite{wang2003multiscale}, FSIM~\cite{zhang2011fsim}, Gradient magnitude similarity deviation (GMSD)~\cite{xue2013gradient}, and visual saliency-induced index (VSI)~\cite{zhang2014vsi}.

    \item Information theoretic estimation: These metrics measure the approximation of the mutual information between the perceived reference and distorted images. Here, we used visual information fidelity (VIF)~\cite{sheikh2006image} and spatial domain VIF (VIFs) metric that quantifies how much information from the reference image is preserved in the distorted image.
    
    \item Learning-based quality estimation: These approaches learn the quality score by training a set of distorted images and corresponding perceptual distance. In this category, we selected two quality measurements: learned perceptual image patch similarity (LPIPS)~\cite{zhang2018unreasonable} model and deep image structure and texture similarity (DISTS)~\cite{ding2020image}.

\end{itemize}


\begin{table}[tbp]

    \small
	\centering
	\begin{tabular}{lcc}
    \hline	
		\textbf{Category} &\textbf{Metric}    \\ 
		\hline
		\multirow{2}{*}{Error visibility} &MAD \cite{larson2010most}& \\
		& NLPD \cite{laparra2016perceptual}&   \\\hline		
		\multirow{5}{*}{Structural similarity} &SSIM\cite{wang2004image} &  \\
	    & MS-SSIM\cite{wang2003multiscale} &  \\
	    & FSIM\cite{zhang2011fsim}  & \\
		& GMSD\cite{xue2013gradient} &  \\
		& VSI \cite{zhang2014vsi} &  \\\hline
		\multirow{2}{*}{Information theory}& VIF  \cite{sheikh2006image}\\
		&VIFs\cite{sheikh2006image}\\\hline
		\multirow{2}{*}{Learning-based} &LPIPS \cite{zhang2018unreasonable}&\\
		& DISTS \cite{ding2020image}\\
\noalign{\smalllskip} \hline \noalign{\smalllskip}		
	\end{tabular}
	\caption{Table 1. The used qualitative quality assessment metrics}	
	\label{tab:tmqi}
\end{table}

In this work, we do not consider the earliest energy-based full-reference quality assessment methods, such as mean-squared error and PSNR, which are already shown to be poorly correlated with human perception of super-resolved images.

\section{Experimental Results}
\label{sec:results}

This section describes our super-resolution dataset for ODI, objective and subjective quality assessments, and experimental analysis.

\subsection{Dataset}

\begin{figure}
    \centering
    \includegraphics[width=0.7\columnwidth]{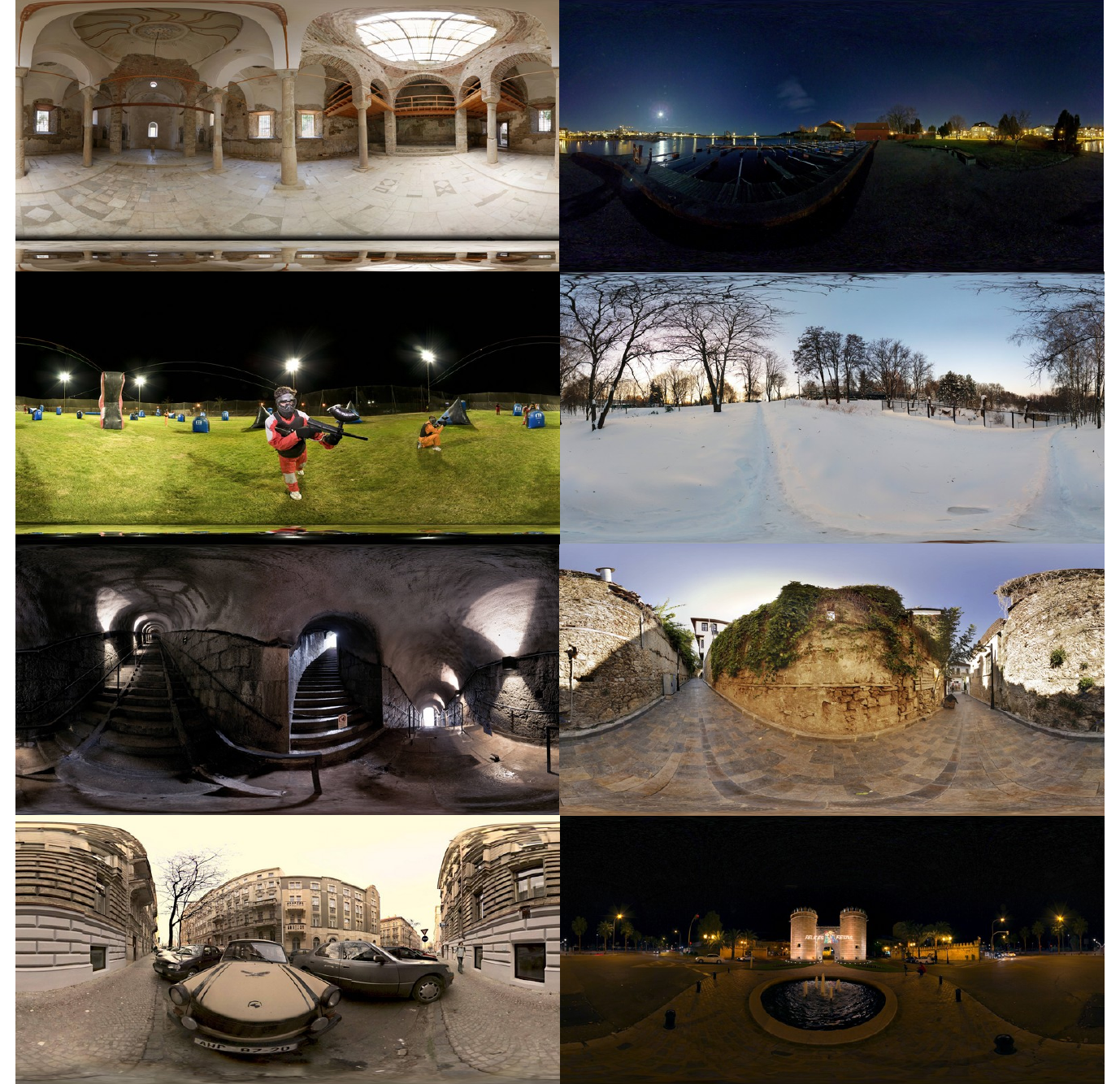}
    \caption{Dataset sample. The selected ODIs from the SUN 360 Panorama Database~\cite{xiao2012recognizing}.}
    \label{fig:dataset}
     \vspace{-1.5em}
\end{figure}

We selected a total of eight ODIs with a diverse content characteristic from the SUN 360 Panorama Database for objective and subjective testing
~\cite{xiao2012recognizing}. This database is the only publicly available ODI dataset with many high-resolution 360-degree images with a wide variety of content, including indoor, outdoor, structures, and human faces. The selected eight ODIs represents four different categories: Structures, Daytime, Indoors, Night-time, Textures and Humans.  Each ODI is in ERP format with a resolution size of 1920$\times$960. The selected ODIs are visualized in Fig.~\ref{fig:dataset}.

We built our SISR dataset for the selected ODIs using two different approaches: CNN-based and GAN-based. For CNN-based methods, we used two state-of-the-art SISR methods: EDSR~\cite{lim2017enhanced} and WDSR~\cite{yu2018wide}. For GAN-based methods, we fine-tuned EDSR and WDSR with the discriminator of the SRGAN~\cite{ledig2017photo}. Hence, we analyzed the following SISR methods for ODI using our proposed quality assessment framework: EDSR-CNN, EDSR-GAN, WDSR-CNN, WDSR-GAN.


For training of EDSR-CNN and WDSR-CNN, we used the default settings of the two SISR~\cite{lim2017enhanced, yu2018wide} methods with $\times$4 scaling factor. Our experiments were carried out on $\times$4 scaling factor between LR and HR. We obtained the LR images by down-sampling HR images with bicubic interpolation. We used DIV2K dataset~\cite{div2k} during the training. Data augmentation techniques such as random cropping, flipping, and rotation were applied to scale up the training dataset. To train each model, we set the batch size to $16$. The weights of all the layers were initialized randomly, and the network was trained from scratch. Further, to optimize the network, we used the ADAM solver with initial learning rates of $1e-4$ and $1e-3$ for EDSR and WDSR, respectively. We fine-tuned EDSR-CNN and WDSR-CNN using the discriminator of the SRGAN~\cite{ledig2017photo}.

\subsection{Objective quality assessment}

We performed objective quality assessment experiments using the proposed objective quality assessment framework. In this experiment, we utilized eleven image quality metrics as described in Section~\ref{proposed_tang}: Input Tangential Views. For this, we used the PyTorch implementations released by Keyan et al.~\cite{ding2020comparison}.

Table~\ref{obj_results} shows the objective quality results using the proposed quality assessment framework.
Objective quality results reveal that majority of the quality metrics shows higher results for CNN-based models (EDSR-CNN and WDSR-CNN) compared to GAN-based models (EDSR-GAN and WDSR-GAN). Interestingly, two quality metrics, NLPD and GMSD show the superiority of the EDSR-GAN model. The reason could be that these two metrics take the human visual system into account and are highly accurate for predicting perceptual quality on noisy images.

\begin{table}[tbp]
\resizebox{0.99\linewidth}{!}{
\begin{tabular}{c c c c l}
\hline	
\multicolumn{4}{c}{\textbf{SISR Models}} &\multirow{2}{*}{\textbf{Metric}}\\
EDSR-CNN &WDSR-CNN &EDSR-GAN &WDSR-GAN\\
\hline
\textbf{93.05} &93.67 &113.16 &107.62 &MAD $\downarrow$\\
0.21 &0.21 &\textbf{0.33} &0.29 &NLPD
$\uparrow$\\
\textbf{0.87} &\textbf{0.87} &0.69 &0.73 &SSIM $\uparrow$\\
\textbf{0.95} &\textbf{0.95} &0.87 &0.87 &MS-SSIM $\uparrow$\\
\textbf{0.93} &\textbf{0.93} &0.88 &0.89 &FSIM $\uparrow$\\
0.09 &0.09 &\textbf{0.12} &0.11 &GMSD $\uparrow$\\
\textbf{0.98} &\textbf{0.98} &0.96 &0.97 &VSI 
$\uparrow$\\
\textbf{0.35} &\textbf{0.35} &0.11 &0.13 &VIF 
$\uparrow$\\
\textbf{0.32} &\textbf{0.32} &0.19 &0.22 &VIFs $\uparrow$\\
\textbf{0.29} &\textbf{0.29} &0.37 &0.36 &LPIPS $\downarrow$\\
\textbf{0.13} &0.14 &0.15 &0.15 &DISTS $\downarrow$\\
\noalign{\smalllskip} \hline \noalign{\smalllskip}	
\end{tabular}
}
\caption{Table 2. Objective metric results of the proposed quality assessment framework. Here, $\uparrow$ means
the higher the better, while $\downarrow$ means the lower the better. Best are marked in \textbf{bold}.}
\label{obj_results}
\end{table}

\subsection{Subjective quality assessment}

We performed a subjective pairwise comparison to validate the perceived quality of different super-resolved ODIs generated by the state-of-the-art methods. In particular, we are interested to investigate CNN-based versus GAN-based SISR results. 

\subsubsection{Test Environment and Setup}

The study was performed in a crowd-sourced format by showcasing a video-like representation of the ODIs. A total of 20 people participated in this subjective study, with age range of 23-40 years, normal or corrected-to-normal vision. We generated a viewport video with 10 seconds for each ODI using the 360Lib software \cite{ye2017algorithm}. The viewport was generated by using the rectilinear projection along with the given field-of-view information. Dynamic viewport setting was selected to create a video-like viewport representation for each ODI.


Participants were able to access the experiments using their web-browser where two super-resolved ODIs corresponding to a scene is displayed for 10 seconds. A total of 48 pairwise combinations were generated from eight scenes and their four super-resolved images obtained using EDSR-CNN, EDSR-GAN, WDSR-CNN and WDSR-GAN models.

\subsubsection{Procedure and Analysis}
\begin{figure}[!tb]
    \centering
    \includegraphics[width=0.5\columnwidth]{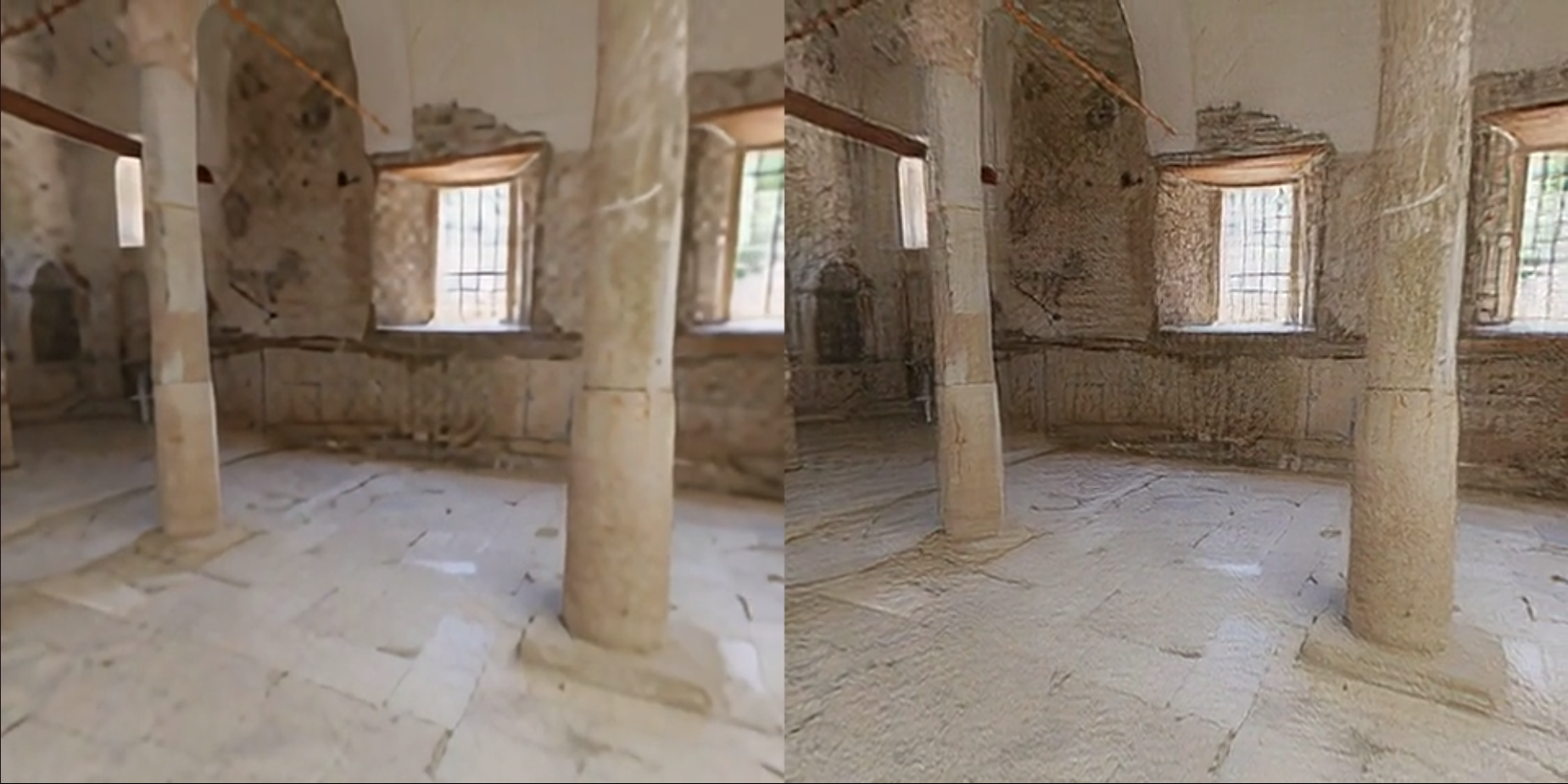}
    \caption{ODI shown side by side during Subjective Test}
    \label{fig:subexp}
     \vspace{-1.5em}
\end{figure}

Procedure: A pair-wise subjective experimental study was performed where each observer was asked to choose an image by showing a pair of images side-by-side. We excluded the option of showing the same image type to force the users to choose one of the stimuli. Each participant was asked to select an image which is more natural and realistic-looking. Participants were provided with unlimited time to make their decision and record their selection. The experiment was divided into a training and test session, where training involved each participant being briefed to familiarize with the subjective quality evaluation task. Each observer compared a pair of image twice, having each super-resolved image displayed on both sides. Figure~\ref{fig:subexp} shows the subjective experiment setup shown to participants.

Analysis: We compute the statistical analysis of pairwise subjective tests. To this end, we scaled the winning frequencies of each SISR model to the continuous quality-scores using the widely known Bradley-Terry (BT) model. The scaling is performed using the statistical analysis proposed in to determine whether the perceived visual quality difference of the compared models is statistically significant. The preference probability Pref-Prob is mathematically given as: 

\begin{equation}\label{eqn}
  Pref-Prob_{m} = \frac{w_m}{n} + \frac{\tau}{2n},
\end{equation} 

where $w$ is the winning probability of the method $m$. $n$ is the total number of participants and $\tau$ is the tie frequency. Given the statistical basis of the subjective study, each method in the pairwise comparison shares equal probability of occurrence (\ie $0.5$) when compared with any other method and hence, the analysis follows a binomial distribution. Based on different state of the art subjective studies to establish initial hypothesis, we followed and performed a Binomial test on the set of collected data and the critical thresholds were determined by plotting the cumulative distribution function of Binomial distribution. We set $95\%$ as the upper level of significance, if we receive $13(B(13,20,0.5) = 0.9423)$ or more preference votes for  a given method over another competing method,  we  consider the given method to be significantly favored in terms of subjective quality with more natural appeal. Similarly, we set $5\%$ as  the lower significance  level,  if we  receive  $6(B(6,20,0.5) = 0.0577)$ or less votes for a given method,  we  consider the given method for an ODI to  be least favored in terms of subjective quality.

\begin{figure}[!tb]
    \centering
    \includegraphics[width=1\columnwidth]{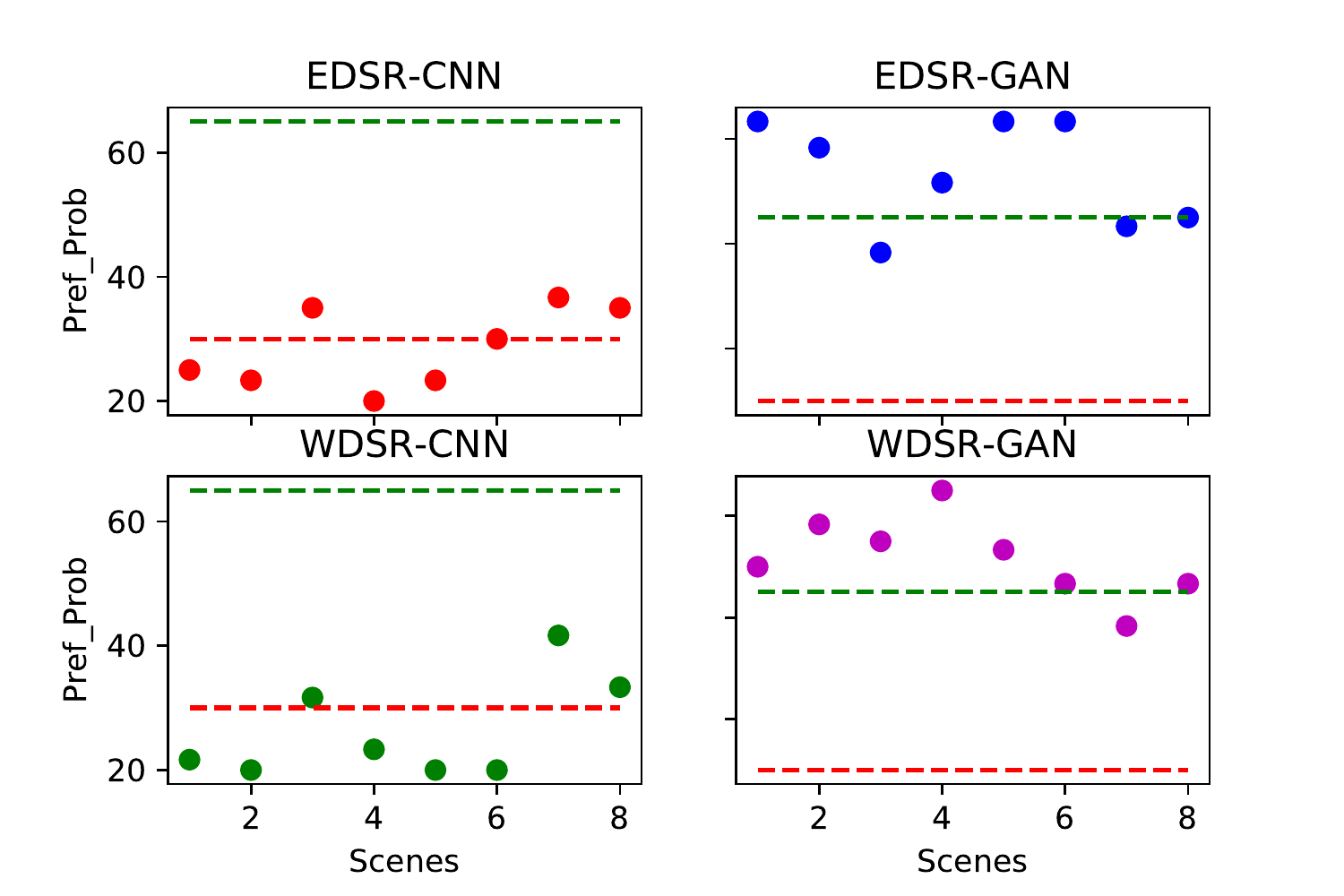}
    \caption{Subjective Test Results.  Preference  probability  of  four techniques (EDSR-CNN, WDSR-CNN, EDSR-GAN, WDSR-GAN) for 8 ODI scenes.}
     \vspace{-0.5em}
    \label{fig:SubTest}
\end{figure}

The subjective tests results for four different methods are shown in Figure \ref{fig:SubTest}. The green and red line mark probabilities of high $(13/20 =.65)$  and low $(6/20 =.30)$ favor-abilities respectively. Our subjective tests results demonstrate that images generated with GANs are more perceptually appealing for different types of scenes compared to CNNs based models.

\begin{table}[tbp]
\resizebox{0.99\linewidth}{!}{
\begin{tabular}{c c c c l}
\hline	
\multicolumn{4}{c}{\textbf{SISR Models}} &\multirow{2}{*}{\textbf{Metric}}\\
EDSR-CNN &WDSR-CNN & EDSR-GAN &WDSR-GAN\\
\hline
4.76 &11.90 &50 &33.33 &MAD $\downarrow$\\
7.14 &7.14 &50 &33.33 &NLPD $\uparrow$\\
35.71 &38.09 &0 &16.67 &SSIM $\uparrow$\\
42.85 &35.71 &0 &16.67 &MS-SSIM $\uparrow$\\
38.09 &35.71 &2.38 &11.90 &FSIM $\uparrow$\\
2.38 &9.52 &50 &33.33 &GMSD $\uparrow$\\
35.71 &33.33 &0 &16.67 &VSI $\uparrow$\\
42.85 &40.48 &0 &16.67 &VIF $\uparrow$\\
38.09 &35.71 &0 &16.67 &VIFs$\uparrow$\\
7.14 &7.14 &42.86 &40.48 &LPIPS$\downarrow$\\
14.28 &23.81 &33.33 &26.19 &DISTS $\downarrow$\\\hline
14.27 & 13.23 & 36.67 & 35.83 &Sub. Score\\
\noalign{\smalllskip} \hline \noalign{\smalllskip}	
\end{tabular}
}
\caption{Table.3 Comparisons of Preference Probabilities of the objective and subjective studies computed over 8 scenes.}\label{obj_sub_results}
\end{table}
        
\subsection{Comparison}
We showcase the comparisons between objective and subjective analysis in Table~\ref{obj_sub_results} using the average preference probability of EDSR-CNN, EDSR-GAN, WDSR-CNN, WDSR-GAN models. These scores are computed using the equation \ref{eqn} by performing a pairwise comparison between objective scores of 4 modalities of a given scene and metric. All 8 scenes that are used in subjective experiments are utilized to obtain the average probability scores. For metrics where higher scores signify better performance, models with higher average probability scores are better and vice versa (shown with up and down arrows).

\begin{figure}[!tb]
    \centering
    \includegraphics[width=1\columnwidth]{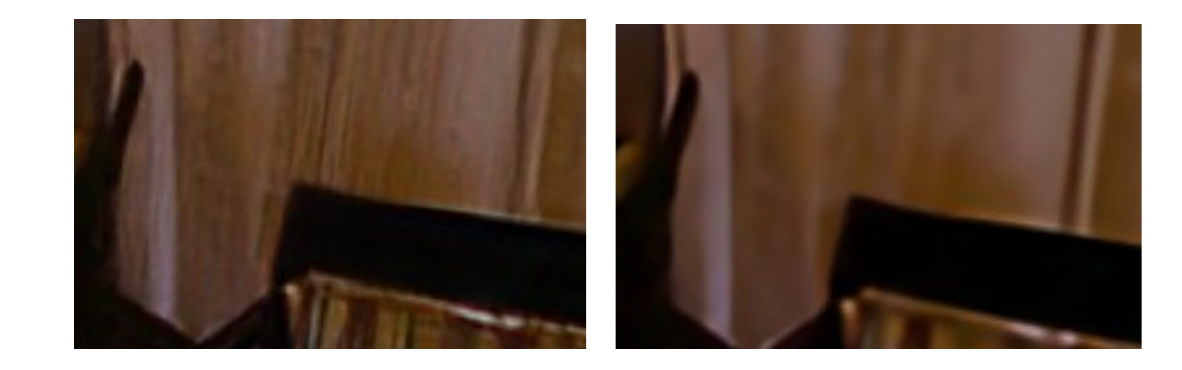}
    \caption{Visual comparison between EDSR-GAN (left) and EDSR-CNN (right).}
    \label{fig:viz_result}
    \vspace{-1.5em}
\end{figure}

While GAN models show superior performance in subjective tests over conventional CNN architectures, all the objective metrics clearly favor the CNN models except the GMSD and NLPD metrics. GANs super-resolved output contains sharp details compared to blurred out CNN SISR network output, which appears realistic perceptually. Such sharp details are however contains visually-imperceptible noise, which hampers objective metric performance. Fig~\ref{fig:viz_result} shows the comparison between GAN-based (EDSR-GAN) and CNN-based (EDSR-CNN) SISR outputs. 

GMSD and NLPD metric show high correlation with subjective analysis and hold potential for designing optimal SISR models for ODIs in future. 
Other metrics, however, appear to be sub-optimal to be used for quality assessment of super-resolved ODIs and need further introspection.

\section{Conclusion}
\label{sec:conclusion}
We presented an objective quality measurement framework which is able to measure super-resolution quality on tangential views. Distortion-free tangential views helps in comparing the subjective analysis and the results from objective quality metrics that are originally designed for 2D image compression methods. In addition, we investigated CNN-based and GAN-based super-resolution models in the objective and subjective experiments. Our analysis showcased how GAN-based super-resolution models show superior performance in subjective tests over conventional CNN architectures. Also, we found that all the objective metrics clearly favor the CNN models except the GMSD and NLPD metric. Our future work will focus on designing a robust SISR models for ODIs.
\footnotesize
\bibliography{ms}
\bibliographystyle{IEEEtran}

\end{document}